# Solar wind triggering of geomagnetic disturbances and strong (M>6.8) earthquakes during the November – December 2004 period

**G. Anagnostopoulos, A. Papandreou and P. Antoniou**

Demokritos University of Thrace, Space Research Laboratory, 67100 Xanthi, Greece

**Abstract.** This paper brings space weather prediction close to earthquake (EQ) prediction research. The results of this paper support conclusions of previously presented statistical studies that solar activity influences the seismic activity, this influence is mediated through rapid geomagnetic disturbances and the geomagnetic disturbances are related with increases of solar wind speed. Our study concern an example of 40 days with direct response of a series of 7 strong-to-giant (M=6.8-9.3) EQs (including the Andaman-Sumatra EQ) to solar wind speed increases and subsequent geomagnetic fast disturbances. Our analysis for 10 M>6 EQs from November 23 to December 28, 2004 suggests a mean time response delay of EQs to fast geomagnetic disturbances of ~1.5 days. The two giant EQs during this period occurred after the two fastest geomagnetic variations, as revealed by the ratio of the daily Kp index variation over a day ΔKp/Δt (12 and 15, respectively). It suggests that the fast disturbance of the magnetosphere, as a result of the solar wind speed increase, is a key parameter in a related space weather-earthquake prediction research. The Solar-magnetosphere-lithosphere coupling and their possible special characteristics during the period examined needs further investigation, since it could provide significant information on the underlying physical relation processes of strong earthquakes.

# 1. Introduction.

The presence of Sun controls to a highest degree the planet Earth, the life on it and, of course, human life, health and culture. In particular, scientific research in the recent decades has greatly improves our knowledge of Sun, its emissions and several aspects of its influence on other planets and their magnetospheres and, most of all, on earth's magnetosphere. In addition, the last two decades two different sciences, space physics and seismology, have been developed to a common new research direction: space whether prediction and earthquake prediction. Space whether prediction research has been accepted as the main priority in space physics, whereas earthquake prediction is a rather controversial issue, despite increasing evidence from the investigation of various physical phenomena about the existence of various types of earthquake precursory signals. In another accompanying paper of this issue we review some of the various methods which provide earthquake precursory signals and we suggest a new one, based on the space observation of energetic electron precipitation in the upper ionosphere (Anagnostopoulos et al.).

The question addressed in this paper is the possible synergy between space whether prediction and earthquake prediction research. Gousheva et al.(2003) noted that although many authors have studied the role of extraterrestrial factors in terrestrial seismicity, the problem remains controversial. However, there is an increasing evidence for an influence of solar activity upon seismic activities. For instance, Khain and Khalilov (2008) presented statistical results that (a) in both the spectra of earthquakes with M>= 7 and Wolf numbers, the main harmonic was found at T ≈ 10-11 years and (b) the most stable components of the spectra of time series of earthquakes and volcanoes are the harmonics at T ≈ 10-13 years and T ≈ 22-24 years. The authors suggested a long-term forecast for seismic activity until 2018 based on its high correlation with solar activity. Odintsov et al. (2007) determined that the maximum of seismic energy released from EQ sources during an 11-yr of the solar sunspots is observed during the phase of solar cycle decline and it lags behind 2yr of the solar maximum.

Most importantly, a relation between the solar wind speed, as it incidents on the Earth's magnetosphere and the seismic activity has been established. It is known that the main source of high speed solar streams are the solar coronal mass ejections (CMEs), which have a maximum near to the solar sunspot maximum as a consequence of the increase solar activity, and the coronal holes with a maximum on the ascending and descending phase of solar activity. Gousheva et al. (2003) performed a statistical study and they found two maxima in the global yearly number of EQs in the 11-yr sunspot cycle, one coinciding with the solar sunspot cycle maximum, and the other one on the descending phase of the solar cycle. In agreement with these results, Odinstov et al (2007) confirmed that the maximum of the earthquakes number directly correlates with a sudden increase in the solar wind velocity.

A significant increase in the incident solar wind speed on the earth's magnetosphere produces magnetospheric storms and substorms. Several scientists searched for a relationship between geomagnetic storms and tectonic events and implemented a transformation of solar flare energy into the energy of tectonic processes. For instance, Sobolev *et al.*(2001) and Zakrzhevskaya and Sobolev (2002) examined the probable seismicity effect of the magnetic storms by analysing the regional seismic catalogue of Kazakhstan and Kyrgyzstan (Mikhailova, 1990). They come to the conclusion that there is a good correlation between magnetic storms and the seismicity, and that the seismic response to a magnetic storm lasts for 10 days, with a maximum effect between 2 to 7 days after the sudden commencement. These authors found a correlation between magnetic storms / sudden commencement and earthquakes in the seismically active region of Kazakhstan and Kyrgyzstan. They formed two sub-catalogs with events preceding and following magnetic storms and they used epoch superposition methods to show that, at a high significance level, the number of earthquakes occurring after storms increases in some areas and decreases in other, and that the tendency toward an increase of the number of earthquakes after the storm characterizes the region as a whole. A possible interpretation of this fact is the compensation of the storm effect time intervals. If the seismicity in some area increases after a storm, it decreases in the passive period (between storms) and *vice versa*. Let storms cause an increase in seismicity (the positive effect) in an area. This leads, on the one hand, to the energy outflow from other areas where the negative effect is observed and, on the other hand, to a decrease in seismicity in this area in the passive period following the active one. They applied parametric and non-parametric criteria to check the absence or presence of significant distinctions between pre- and post-storm samples something that concluded in the significant difference between events preceding and following magnetic storms (with a probability of 99.9% in the studied areas). They inferred that the storm effect on seismicity is evidently non-accidental.

Bakhmutov et al. (2007) studied the correlation between geomagnetic field disturbances and earthquakes on the seismic zone in Vrancea, South Carpathian region. They demonstrated a reaction of the stress-strained medium to geomagnetic disturbances (morphological features in the geomagnetic variation structure) suggesting that most of the earthquakes were a result of stress discharges.

Furthermore, Tarasov *et al.* (1999) reported a probable triggering effect on anthropogenic impulsive electrical signals on seismicity. Analyzing variations of the number of earthquakes in Kyrgyzstan associated with electrical signals radiated by an MHD generator, they inferred that the number of earthquakes tends to increase 3-4 days after the electrical signal passage. The duration of this activation stage is a few days, and the dimensions of the possibly affected area are on the order of a few hundred kilometres.

In conclusion, the above referred papers based on statistical results are consistent with an increase of seismicity of earth under influence of geomagnetic storms related with the solar wind speed increases. In this paper we study for the first time the possible direct influence of high solar wind streams on the earth's seismicity for a time interval of 35 days including the Sumatra – Andaman strong (9.3) EQ (Stein and Okal, 2007), that caused the tsunami of hundreds thousands dead people. Our data analysis clearly shows that 7 of the 7 solar wind increases observed upstream from the earth's bow shock from 23/11/2004 to 28 /12/2004 were followed by one or more M>6 earthquakes (M=7.1, 7.0, 6.8, 6.8, 8.1, 9.3, respectively) within ~1.5 days, all (except the Sumatra EQ) occurred at the edges of Pacific Plate. The possible conditions of the solar wind-magnetosphere-lithosphere coupling during that special time period are also discussed.

# 2. Observations

## 2.1 Data and Processing

In the present study, the geomagnetic Kp index and the Solar Wind velocity (Kp and SwV) data have been retrieved from the Space weather prediction center (http://www.swpc.noaa.gov/) and the earthquake data from http://earthquake.usgs.gov. The relevant time period from 23/11/2004 to 31/12/2004 is examined. We have plotted these data in composite figures in an attempt to search for any qualitative correlations between parameters of space weather, geomagnetic status and earthquakes. Additionally we present in tables the quantitative results in order to provide a first quantification of these correlations.

## 2.2 Data Analysis.

Figure 1 shows an overview of the magnitude of strong (M>6) earthquakes (top panel), as well as time series of (3-hours) geomagnetic Kp index (middle panel) and 3-hours averaged solar wind speed (bottom panel) for time interval November 23 to December 28, 2004, which includes the giant Andaman-Sumatra EQ. This figure shows in brief the main characteristics of the solar -magnetosphere-lithosphere coupling discussed in this paper during this exceptional case of seismic activity. The green normal lines mark times of solar wind speed increases, the blue arrows indicate related increases in the daily number of the Kp index, and the red arrows point the EQs following the corresponding solar wind and Kp index values increases. From this first figure it is clearly seen an association between solar wind and Kp index values increases with occurrence of strong EQs. In the following figures this relation is examined in detail.

Figures 2 and 3 display the Kp and SwV along with their variations in different time scales (daily and 3hourly respectively). In Figure 4 we present the earthquake activity along with the aforementioned space weather parameters. As can be seen, in the temporal vicinity of every major earthquake (M>6 in the Richter scale) there are local maxima in the Kp index. In all of the cases these maxima of the Kp occur exactly before or the same day of the earthquake. In addition to Figure 4 we present Tables 1 and 2 in which we have evaluated the slopes of the Kp up to the maximum that is relevant to each earthquake.

It is interesting to note that although for some earthquakes the daily Kp increase seems rather gradual (low Kp slope), this is a result of averaging. This is especially evident in earthquakes #6 and # 7 for which we have calculated the Kp slope for all the monotonously increasing Kp dates and then calculated the slopes of some actual 'day to day' Kp variations. So, in the case of earthquake #6, while the total calculated Kp slope is 6, it increases to 8 when calculated for two days prior to the earthquake (Kp=13 and 21 respectively). This is even more accented in earthquake #7 in which the Kp slope is 8 and 9, when calculated from 4 and 3 points respectively, while from 'day 3' before the earthquake (20/12/2004) to 'day 2' before the earthquake (21/12/2004), the actual Kp slope is 12.

In addition to these results we present in figures 4 and 6 the temporal distribution of the Kp index and the solar wind speed (SwV) maxima regarding each earthquake. While the solar wind activity is not definitive in the vicinity of each earthquake, the Kp activity appears to presents a temporal association to the occurrence of earthquakes. To verify this association, we have tabulated the earthquake events and recorded the temporal distance (in 3 hourly periods) of each Kp maximum that occurred before the earthquake (Table 3). As is shown in Table 3 there is an average preceding Kp maximum at about 36 ± 3 hours between the Kp maximum and the earthquake occurrence.

Figure 7 shows magnetic field and solar wind data from the ACE spacecraft suggesting that the high speed solar wind streams were due to corotating interaction regions (CIRs) incident on the earth's magnetosphere.

These results although preliminary and mostly qualitative seem to suggest that a possible association between the space weather parameters and geological phenomena could be established, should a larger data sample could be explored. It is clear that such a larger scale study could conceivably lead to a possible correlation of the actual physical processes of space magnetism as a triggering means for geological processes that would lead to observable effects such as earthquakes.

## 3. Discussion and Conclusion

Several studies in the last years have been devoted to the relations between solar activity and its impact on the earth's magnetosphere, ionosphere, atmosphere, human life and health, technological systems and lithosphere, including earthquakes. Space science community has recently concentrated his interest on the so called "space weather" in order to predict geomagnetic and ionospheric disturbances, which have significant impact on electrical power systems, telecommunications, oil pipelines, spacecraft and aircraft electronics, astronauts safety, climate changes etc. The results of this paper further support the results of previously presented statistical studies that (a) the solar activity influences the seismic activity, (b) this influence is mediated through rapid geomagnetic disturbances, (c) the geomagnetic disturbances is related with the solar wind speed increases. These conclusions brings space whether prediction close to earthquake prediction research.

Various authors attributed seismic activity to a variety of extraterrestrial factors. Different elements of solar activity have been proposed as triggers of seismic activity: solar and lunar tides (Jakubcova and Pick., 1987; Belyakov et al.,2007), solar proton fluxes (Velinov, 1975), earthward movement of the magnetopause caused by an increased solar wind dynamic pressure (Makarova and Shirochkov, 1999), high solar wind speed (Sytinski, 1989). The Sun – interplanetary space – magnetosphere – ionosphere – atmosphere – lithosphere chain is a complicated open dynamic nonlinear system of a complex of processes with high unpredictability. In this context, the seismic phenomena of the Earth should be considered as a part of the whole Sun-Earth system. The relation between seismicity and the Sun-Earth chain processes has been considered as ambiguous. The contribution of various solar impact processes on earth's lithosphere and their association with the seismic energy release needs much work.

How solar wind impact on the magnetosphere can provide energy to lithosphere and trigger seismic EQs? Sobolev et al (2001) and Zakrzhevskaya and Sobolev (2001) argued that the estimation of the energies supplied by magnetic storms and released by earthquakes suggests a trigger mechanism should be considered as responsible for the magnetic storm effect on seismicity. If the crust in the area of the forthcoming earthquake is in metastable state, it becomes more sensitive to trigger-like effects. The electrical energy supplied to the earth during a storm is then converted into mechanical energy via piezoelectric, electrokinetic or other mechanoelectric effects and additionally increases mechanical stresses. Sytinskii (1997) argued that the triggering mechanism for EQ occurrence is the solar induced change in atmospheric circulation expressed in large scale reorganization of |baric fields. Furthermore, Prikyl et al. (2003) provided evidence that the gravity waves are generated by auroral electrojets caused by high speed solar wind MHD waves. Kormiltsev et al. (2002) suggested another mechanism of the effect of magnetic storms on seismicity. They hypothesized that electro-osmotic fluid flow induced by magnetic storms generates anomalous porous pressure, which could be the triggering of the tectonic event.

The data analysis of seven strong (>6.8) EQs between 23/11/2004-28/12/2004, provides a possible useful information for further testing the storm induced triggering mechanism of EQs. These data suggests that strong EQs occurred during the fast increase of the geomagnetic activity as it inferred by the rythme variation of the geomagnetic Kp index ($Kp/\Delta t$) and not during the large storms. Furthermore, the two strongest EQs during this period, the giant Sumatra – Andaman EQ (M=9.3) and north of Macquarie islands EQ (M = 8.1) occurred on December 26,2004 and on December 23,2004, after the recorded highest values of the ratio $Kp/\Delta t$ evaluated for the period of 35 days. Additionally, the six strong EQs before the Sumatra – Andaman EQ occurred on the same broad area of the Pacific |Plate.

This paper provide evidence that the increase of solar wind speed, and the subsequent fast variability of the geomagnetic field was probably a trigger agent for the giant Sumatra – Andaman EQ and for 6 more than 6.8 (>6.8) EQs occurred within the period of a month before that deadly event (Sumatra – Andaman EQ). This hypothesis is strongly supported by the data and is consistent with previously published statistical results mentioned in Section 1.

The solar-magnetosphere-lithosphere coupling and their possible special characteristics during the period examined in this paper should further studied, since it provides a good example of direct response of a series of 7 strong-to-giant EQs to solar wind speed increases and their subsequent geomagnetic disturbances. Moreover, it could provide significant information on the underlying physical relation processes. The whole set of data between 23/11/2004-28/12/2004 and the probably suggestion that the fast variability of the geomagnetic field is a key parameter for formulating a theory on solar wind / geomagnetic storm triggering of the EQs is under examination. Since, it is generally accepted that the strength of the Sumatra – Andaman (SA) EQ was unexpected from only a tectonic plate point of view and that the oceanwide tsunamis may reflect the short earthquake history sampled (Stein and Okal, 2007), we think that the vicinity in time (~3 days) and space (Fig. 8) of the two giant (8.1, 9.3) EQs may suggest a relation of SA EQ with the previous strong EQ which occurred within the broad area where Pacific, Philippine, Eurasian and Indian Plates meets one each other.

**Acknowlegments.** The leading author thanks Drs E. Sarris, G. Pavlos and D. Sarafopoulos for helpful discussions and Dr P. Preka for his valuable comments on the manuscript.

## Figure captions:

Figure 1 An overview of the magnitude of strong (M>6) earthquakes (top panel), as well as time series of (3-hours) geomagnetic Kp index (middle panel) and 3-hours averaged solar wind speed (bottom panel) for time interval November 23 to December 28, 2004. A correlation between solar wind and Kp index values increases with occurrence of strong EQs is evident by arrows

Figure 2 a-d: Kp, ΔKp, SwV, ΔSwV as a function of time (daily time scale)

Figure 3 a-d: Kp, ΔKp, SwV, ΔSwV as a function of time (3 hourly time scale)

Figure 4 a,b,c. Earthquake Magnitudes, Kp and ΔKp in the time interval between 23/11/2004 and 31/12/2004 (daily intervals). Kp reaches a local maximum exactly before or at the day of an earthquake occurrence.

Figure 5 a,b. Earthquake magnitudes and peak Kp variances that occurs before earthquakes.

Figure 6 a,b. Earthquake magnitudes and significant peak SwV variances that occurs before earthquakes.

Figure 7. Solar wind magnetic field and solar wind data suggesting that the high speed solar wind streams were due to corotating interaction regions (CIRs) incident on the earth’s magnetosphere

Figure 8. The sites of strong (>6.8) earthquake epicenters occurred between 23/11/2004 and 31/12/2004.

## Tables

Table 1 : Dates, Kp values and calculated Kp Slopes (ΔKp/Δt) before every earthquake in the time period between 23/11/2004 and 31/12/2004 (daily intervals) .

Table 2: Earthquakes, Dates, magnitudes and Kp Slopes for the time period between 23/11/2004 and 3/1/2005.

Table 3: Table of the time intervals between the time of the max Kp and the occurrence of the earthquake. On average, Kp maximum precedes an earthquake by 36 hours.

**Figure 1**

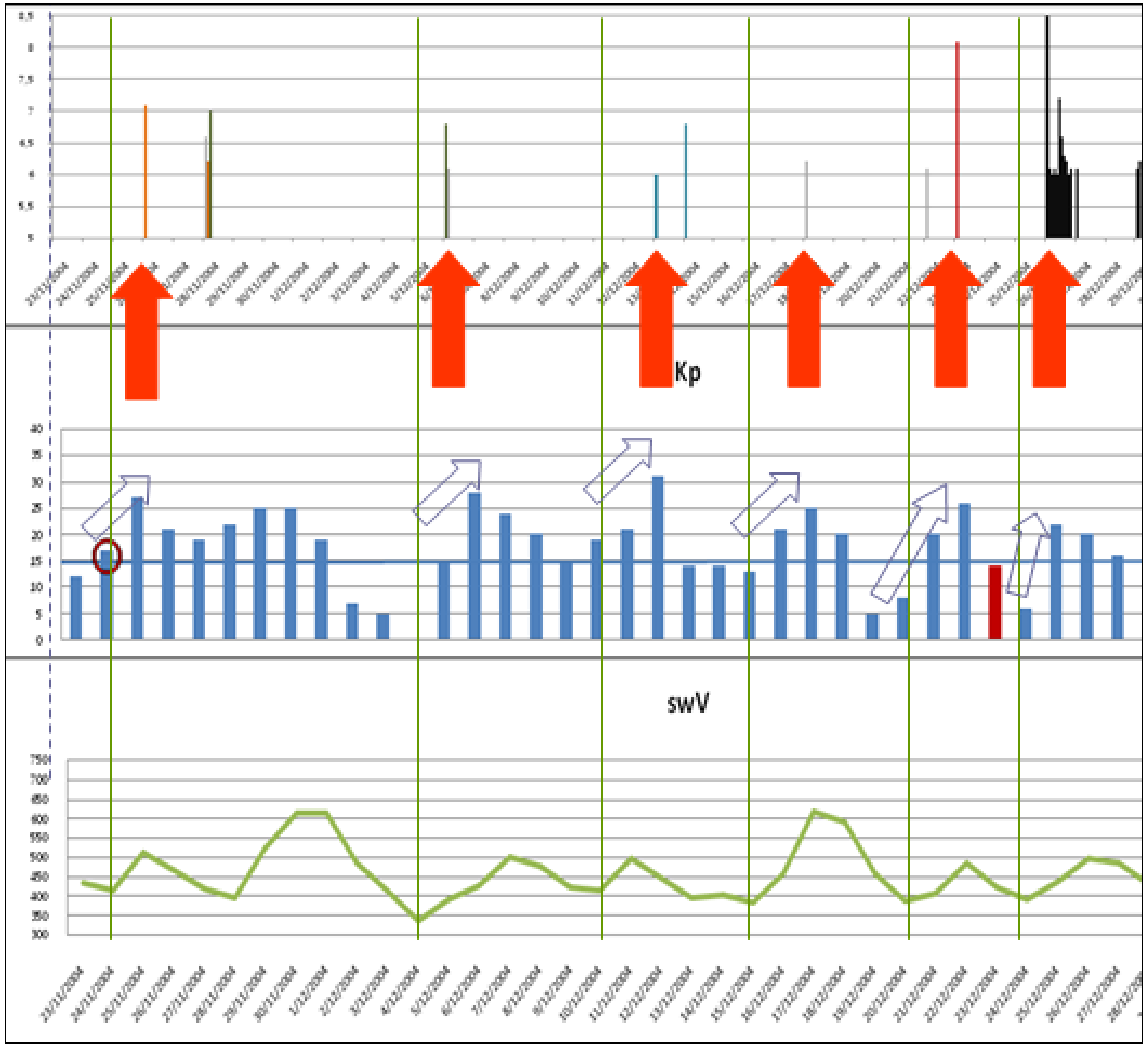

**Figure 2**

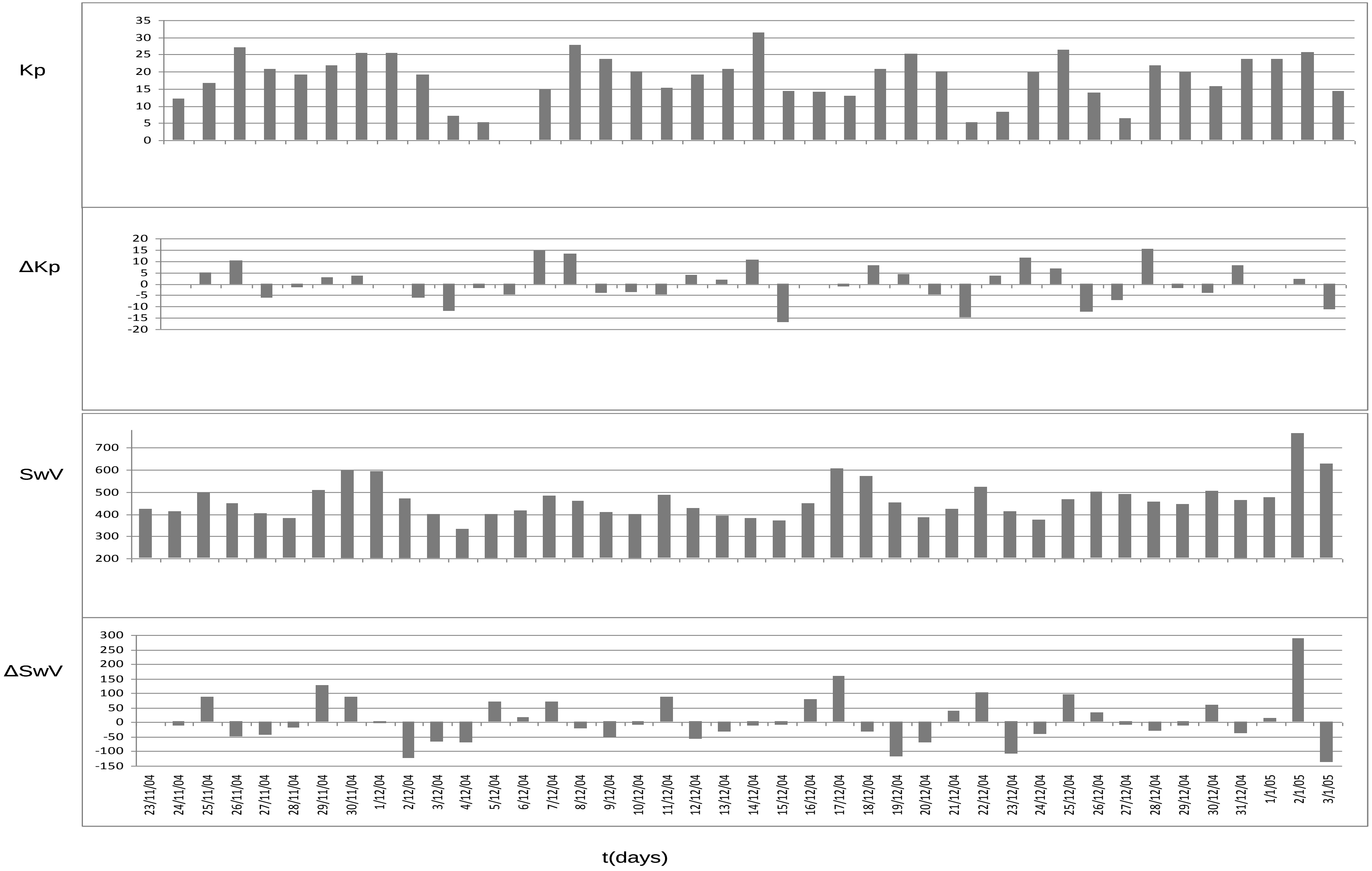

**Figure 3**

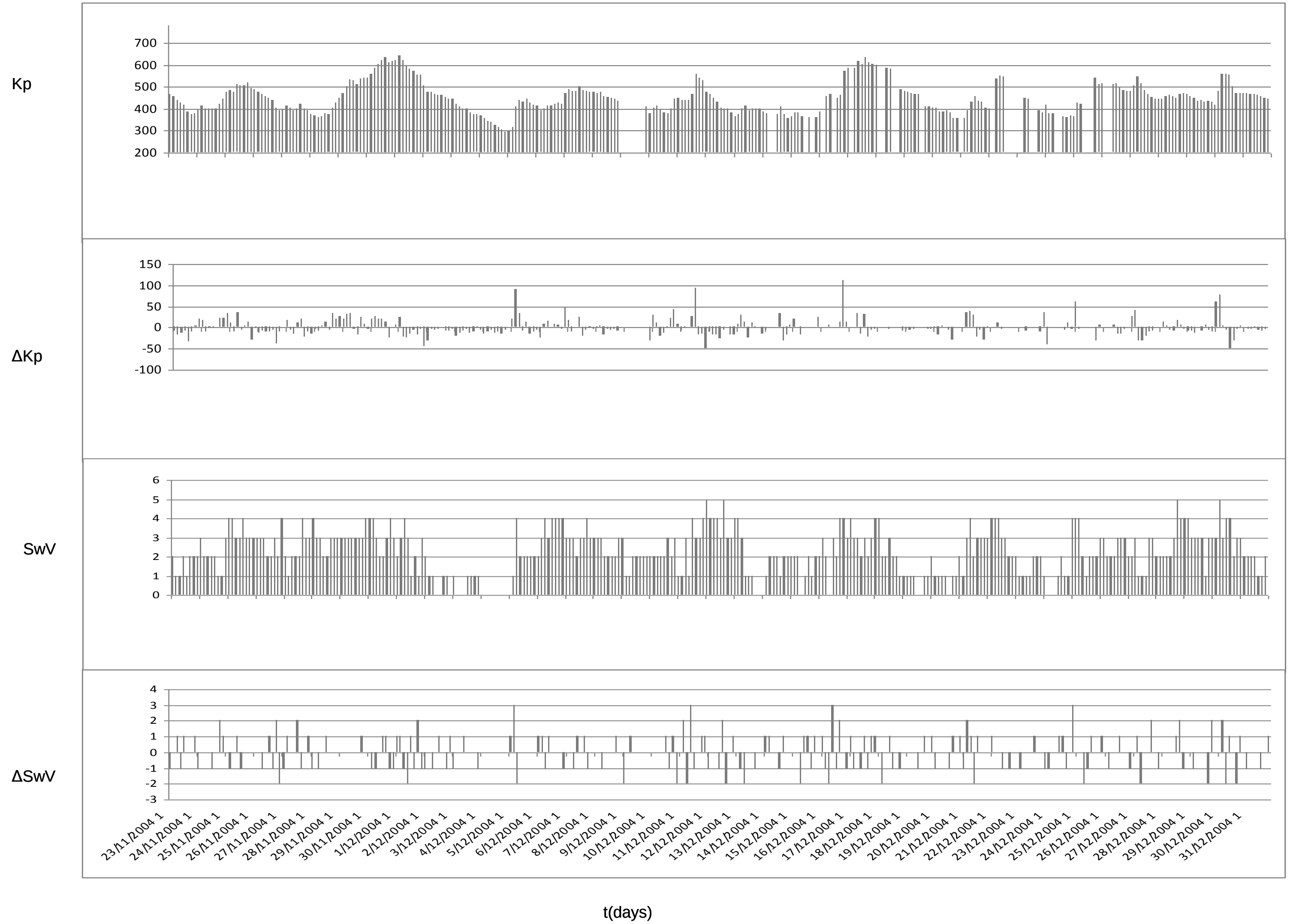

Figure 4

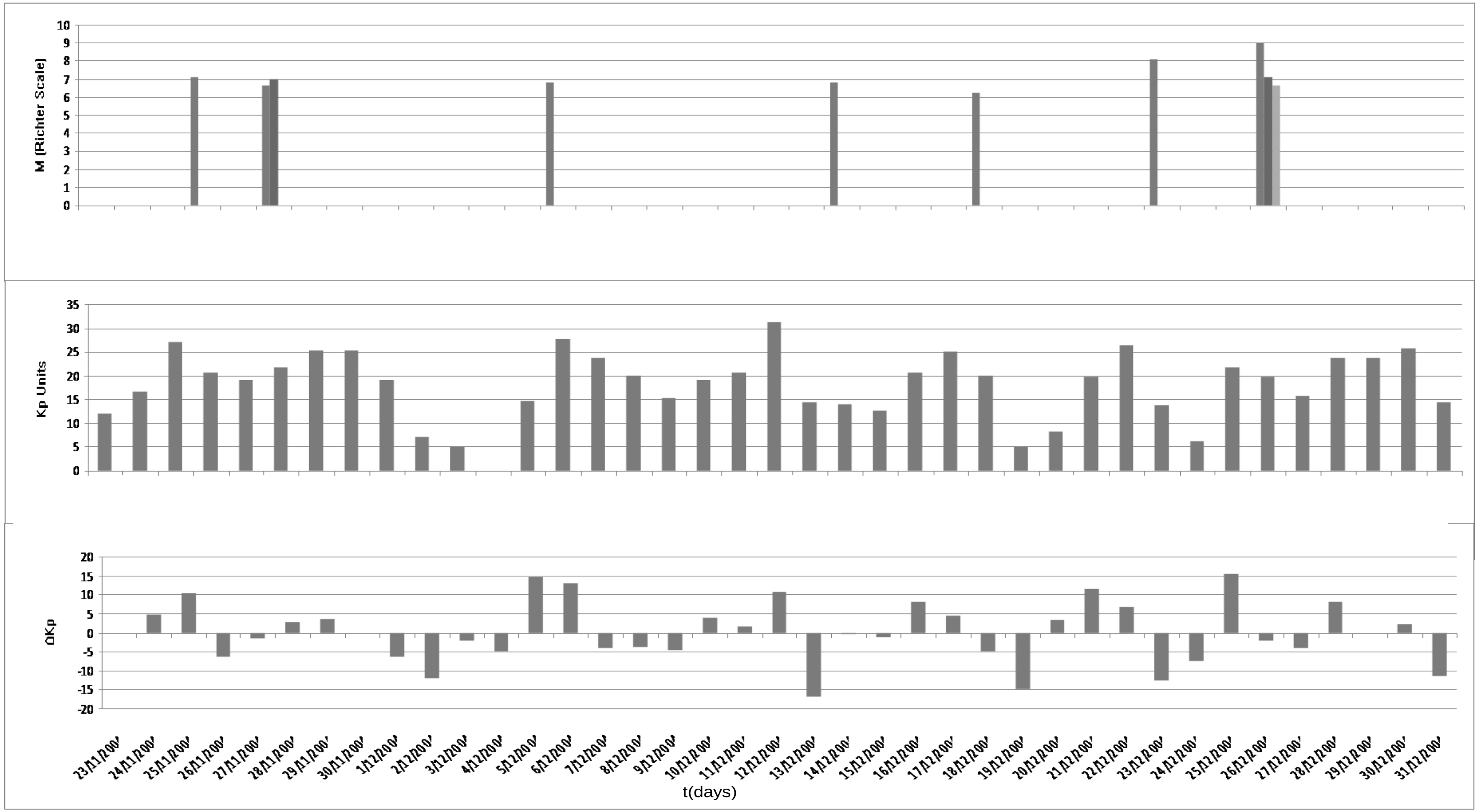

Figure 5

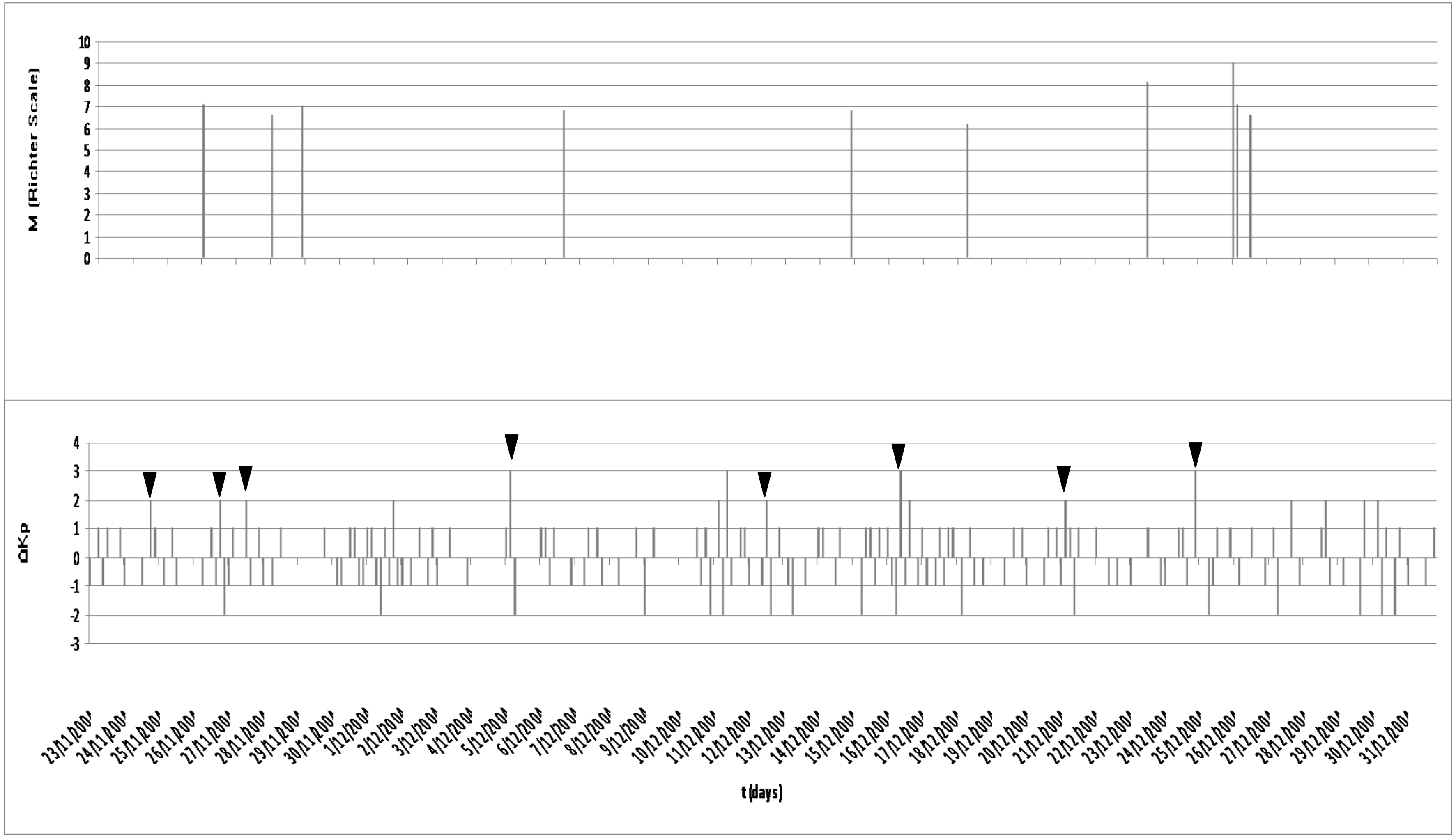

## Figure 6

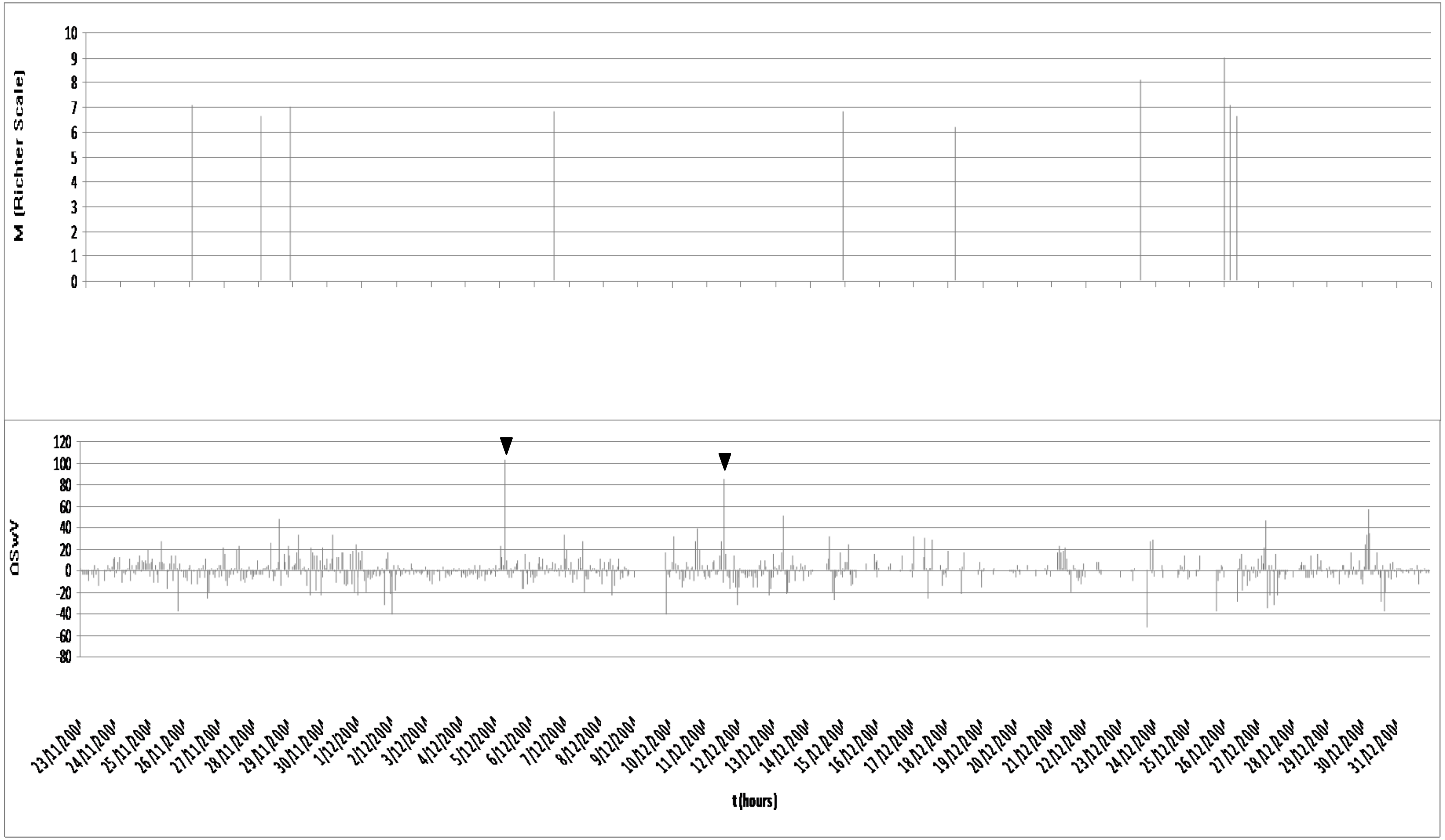

**Figure 8**

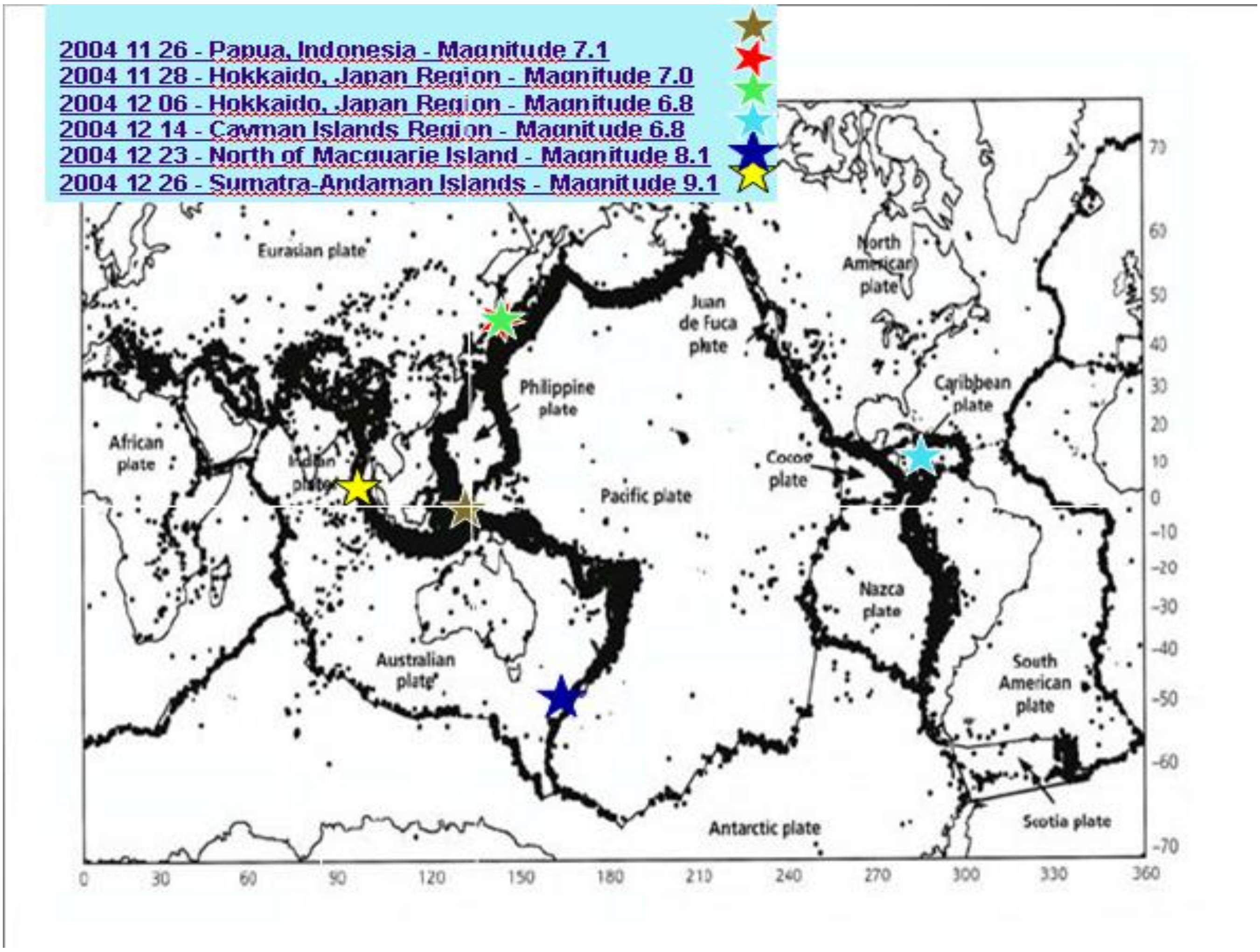

2004 11 26 - Papua, Indonesia - Magnitude 7.1
2004 11 28 - Hokkaido, Japan Region - Magnitude 7.0
2004 12 06 - Hokkaido, Japan Region - Magnitude 6.8
2004 12 14 - Cayman Islands Region - Magnitude 6.8
2004 12 23 - North of Macquarie Island - Magnitude 8.1
2004 12 26 - Sumatra-Andaman Islands - Magnitude 9.1
Eurasian plate
North American plate
Juan de Fuca plate
Philippine plate
Caribbean plate
African plate
Indian plate
Pacific plate
Cocos plate
Nazca plate
Australian plate
South American plate
Antarctic plate
Scotia plate

Figure 7

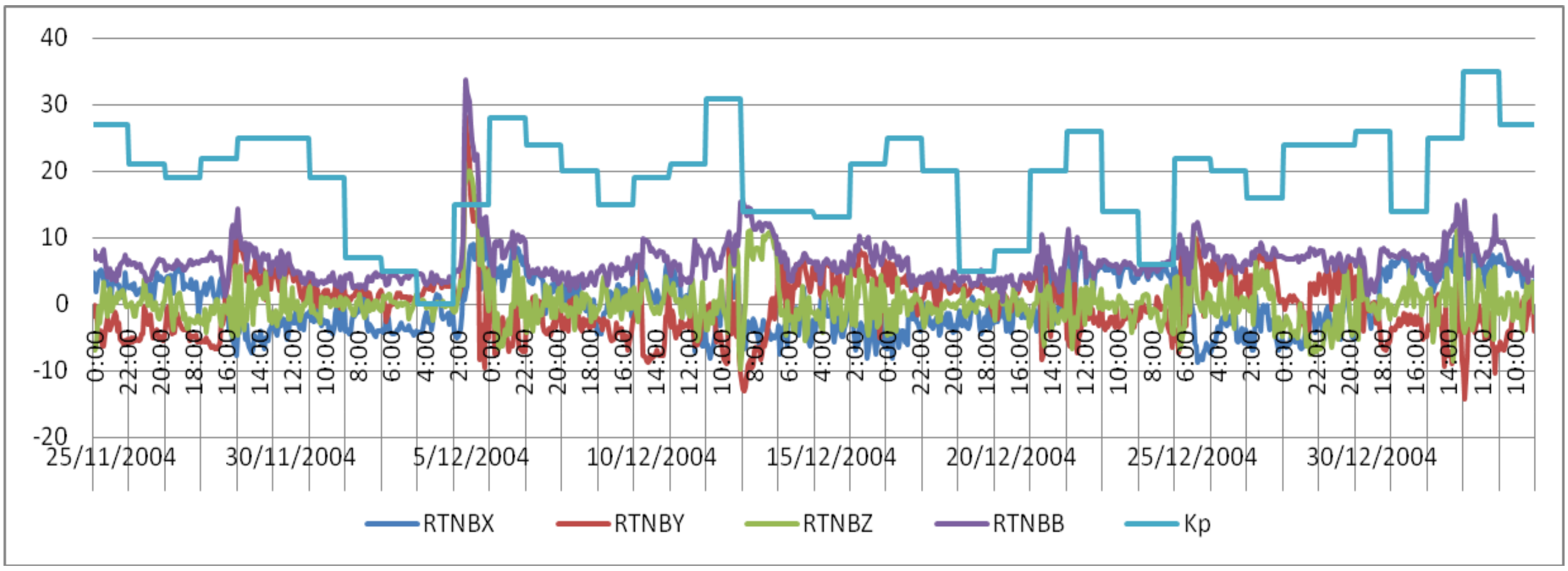

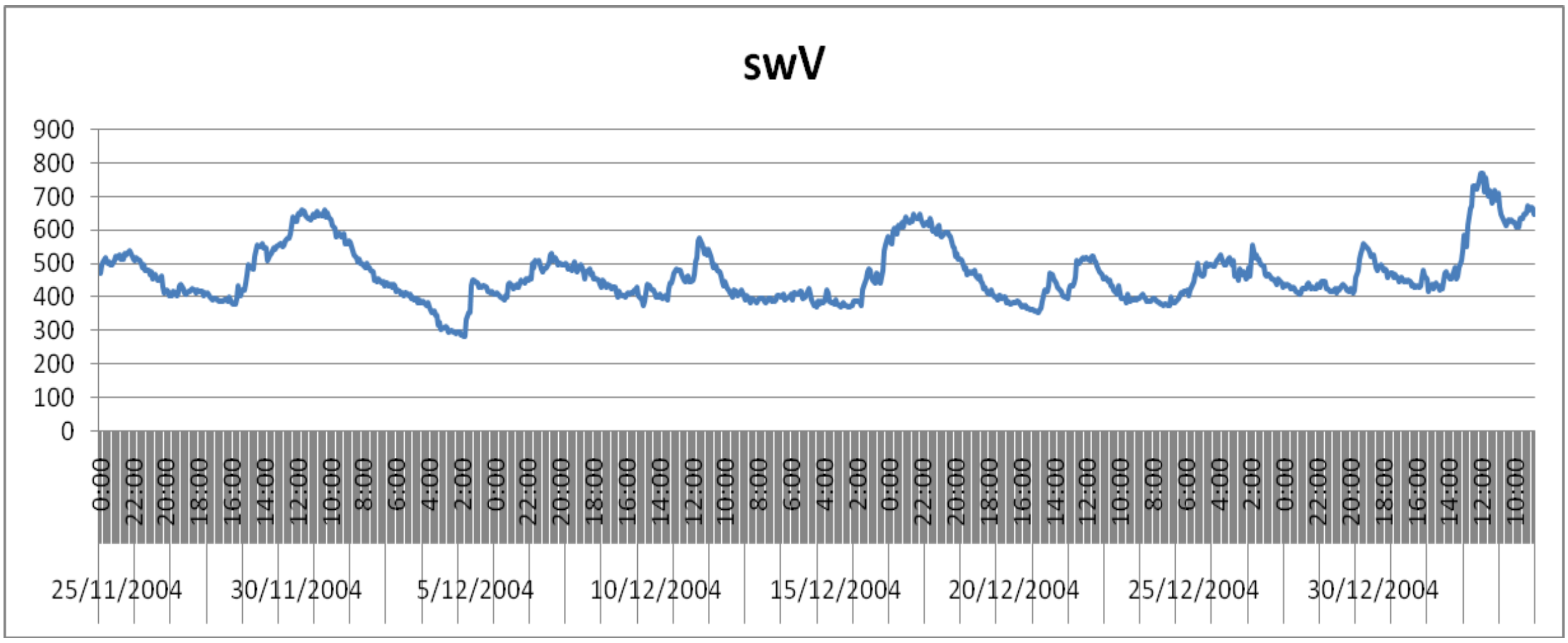

Table 1

| Earthquake No | EQ – Kp Times (Date) | Kp | Kp Slopes (ΔKp/Δt) | ΔKp/Δt time scale (days) |
|---|---|---|---|---|
| 1 | **26/11/2004** | | 8 | |
| | 23/11/2004 | 12 | | |
| | 24/11/2004 | 17 | | |
| | 25/11/2004 | 27 | | |
| 2&3 | **28/11/2004** | | 3 | |
| | 27/11/2004 | 19 | | |
| | 28/11/2004 | 22 | | |
| 4 | **6/12/2004** | | 13 | |
| | 5/12/2004 | 15 | | |
| | 6/12/2004 | 28 | | |
| 5 | **14/12/2004** | | 6 | |
| | 10/12/2004 | 19 | | |
| | 11/12/2004 | 21 | | |
| | 12/12/2004 | 31 | | |
| 6 | **18/12/2004** | | | |
| | 15/12/04 | 13 | 6 | 3 |
| | 16/12/04 | 21 | 8 | 2 |
| | 17/12/04 | 25 | | |
| 7 | **23/12/2004** | | | |
| | 19/12/2004 | 5 | 8 | 4 |
| | 20/12/2004 | 8 | 9 | 3 |
| | 21/12/2004 | 20 | **12** | 2 |
| | 22/12/2004 | 26 | | |
| 8,9,10 | **26/12/2004** | | **15** | |
| | 24/12/2004 | 6 | | |
| | 25/12/2004 | 22 | | |

Table 2

| EQ No | Earthquake (Date) | Earthquake Magnitude | ΔKp/day |
|---|---|---|---|
| 1 | 26/11/2004 | 7,1 | 8 |
| 2&3 | 28/11/2004 | 6,6 - 7 | 3 |
| 4 | 6/12/2004 | 6,8 | 13 |
| 5 | 14/12/2004 | 6,8 | 6 |
| 6 | 18/12/2004 | 6,2 | 6 |
| 7 | 23/12/2004 | 8,1 | 8 |
| 8,9,10 | 26/12/2004 | 9 - 7,1 - 6,6 | 15 |

Table 3

| EQ No | EQ Date | | $t_{EQ} - t_{Max\ Kp}$ (hours) |
|---|---|---|---|
| 1 | 26/11/2004 | 7,1 | 36 |
| 2 | 28/11/2004 | 6,6 | 33 |
| 3 | 28/11/2004 | 7 | 30 |
| 4 | 6/12/2004 | 6,8 | 33 |
| 5 | 14/12/2004 | 6,8 | 33 |
| 6 | 18/12/2004 | 6,2 | 36 |
| 7 | 23/12/2004 | 8,1 | 42 |
| 8 | 26/12/2004 | 9 | 36 |
| 9 | 26/12/2004 | 7,1 | 24 |
| 10 | 26/12/2004 | 6,6 | 39 |
| | | | **Mean 36**<br>SD 3 |